\documentclass[10pt,journal,compsoc]{IEEEtran}

\usepackage{amsmath,amssymb}
\usepackage{graphicx}
\usepackage{booktabs}
\usepackage{array}
\usepackage{xcolor}
\usepackage{wrapfig}
\usepackage{url}
\ifCLASSOPTIONcompsoc
  \usepackage[nocompress]{cite}
\else
  \usepackage{cite}
\fi
\usepackage[ruled,vlined,linesnumbered]{algorithm2e}
\SetAlFnt{\small}
\SetAlCapFnt{\small}
\SetFuncSty{textsc}
\usepackage[hidelinks]{hyperref}

\usepackage{amsthm}

\newtheorem{lemma}{Lemma}
\newtheorem{theorem}{Theorem}


\makeatletter
\def\paragraph{\@startsection{paragraph}{4}{\z@}{1.2ex plus 0.5ex minus 0.2ex}%
  {-0.3em}{\normalfont\normalsize\bfseries}}
\def\@IEEEsectpunct{.\ }
\makeatother

\makeatletter
\providecommand\oddpage@label[2]{}
\providecommand\pgfsyspdfmark[3]{}
\makeatother

\definecolor{AAA}{rgb}{1.0, 0.13, 0.32}
\definecolor{BBB}{rgb}{0.2, 0.1, 1}
\definecolor{CCC}{rgb}{0.0, 1, 0}
\definecolor{DDD}{rgb}{0.0, 0, 1}

\newcommand{\queryCpuOneTableBody}{%
DP-NNS$^1$~\cite{dpnns} & \underline{0.284} & \underline{0.314} & \underline{0.352} & \underline{0.484} & \underline{0.585} & \underline{0.675} & 1.151 \\
DP-NNS$^2$ & 0.314 & 0.461 & 0.465 & 1.145 & 1.656 & 2.168 & 4.369 \\
nanoflann & 1.197 & 2.669 & 1.719 & 6.346 & 8.538 & 6.475 & 3.487 \\
R$^*$-tree & 2.115 & 3.480 & 2.524 & 6.780 & 6.603 & 5.087 & 2.207 \\
Octree & 5.968 & 10.337 & 7.002 & 17.031 & 21.322 & 22.057 & 21.550 \\
BD-tree & 2.830 & 5.978 & 7.422 & 22.034 & 199.028 & 62.443 & 6.750 \\
cKDTree & 4.126 & 15.447 & 16.145 & 51.861 & 158.378 & 139.573 & 25.994 \\
ArborX & 1.057 & 2.026 & 1.247 & 4.923 & 4.688 & 3.295 & \underline{0.817} \\
Ours & \textbf{\underline{0.144}} & \textbf{\underline{0.161}} & \textbf{\underline{0.145}} & \textbf{\underline{0.210}} & \textbf{\underline{0.217}} & \textbf{\underline{0.243}} & \textbf{\underline{0.407}} \\}

\newcommand{\queryCostTableBody}{%
Camel & 190 & 110 & 201 & 116 & 193 & 107 \\
Bunny & 185 & 109 & 211 & 132 & 191 & 108 \\
Dragon & 240 & 104 & 280 & 123 & 235 & 102 \\
Kitten & 226 & 130 & 391 & 194 & 221 & 127 \\
Armadillo & 277 & 112 & 490 & 163 & 276 & 111 \\
Lucy & 332 & 148 & 644 & 226 & 325 & 143 \\
Sponza & 482 & 292 & 958 & 383 & 500 & 300 \\}

\newcommand{\constructionAllTableBody}{%
DP-NNS$^1$~\cite{dpnns} & 147.8 & 368.3 & 515.4 & 1{,}488.7 & 3{,}798.2 & 5{,}524.1 & 6{,}390.5 \\
R$^*$-tree & 3.6 & 9.8 & 14.5 & 44.8 & 128.9 & 187.8 & 248.1 \\
Octree & 1.3 & 4.0 & 3.8 & 13.6 & 36.1 & 42.3 & 117.7 \\
BD-tree & 5.1 & 13.9 & 20.7 & 83.7 & 275.0 & 439.1 & 555.4 \\
cKDTree & 4.1 & 12.0 & 16.4 & 58.7 & 173.0 & 254.0 & 373.8 \\
\multicolumn{8}{@{}c@{}}{\xleaders\hbox to 4pt{\hss\rule[1.1ex]{2pt}{0.4pt}\hss}\hfill\kern0pt} \\[-0.6ex]
nanoflann & 3.3 & 7.3 & 11.0 & 30.4 & 64.3 & 107.4 & 134.0 \\
ArborX & 1.2 & 2.6 & 2.7 & 9.5 & 19.4 & 29.7 & 28.1 \\
DP-NNS$^2$ & 35.2 & 51.3 & 61.2 & 267.0 & 298.2 & 476.4 & 709.2 \\
Ours & 26.6 & 45.6 & 49.4 & 180.5 & 261.2 & 390.2 & 499.0 \\}

\newcommand{\uniformTableBody}{%
DP-NNS~\cite{dpnns} & 0.214 & 0.298 & 0.370 & 0.462 & 0.723 & 0.991 \\
nanoflann & \underline{0.199} & \underline{0.252} & \underline{0.307} & \underline{0.372} & \underline{0.530} & 0.819 \\
R$^*$-tree & 1.108 & 1.335 & 1.415 & 1.501 & 1.708 & 1.942 \\
Octree & 3.179 & 3.853 & 3.827 & 4.215 & 5.341 & 6.315 \\
BD-tree & 0.440 & 0.535 & 0.666 & 0.770 & 1.064 & 1.559 \\
cKDTree & 0.457 & 0.545 & 0.630 & 0.698 & 0.853 & 1.318 \\
ArborX & 0.428 & 0.508 & 0.539 & 0.566 & 0.648 & \underline{0.733} \\
Ours & \textbf{\underline{0.176}} & \textbf{\underline{0.205}} & \textbf{\underline{0.214}} & \textbf{\underline{0.239}} & \textbf{\underline{0.263}} & \textbf{\underline{0.337}} \\}
\newcommand{\gridSortAblationTableBody}{%
$\times$ & $\times$ & 0.280 & 0.304 & 0.331 & 0.480 & 0.586 & 0.663 & 1.148 \\
$\checkmark$ & $\times$ & 0.160 & 0.204 & 0.170 & 0.337 & 0.334 & 0.421 & 0.861 \\
$\times$ & $\checkmark$ & 0.213 & 0.237 & 0.263 & 0.295 & 0.362 & 0.412 & 0.603 \\
$\checkmark$ & $\checkmark$ & \textbf{0.149} & \textbf{0.164} & \textbf{0.147} & \textbf{0.256} & \textbf{0.218} & \textbf{0.247} & \textbf{0.408} \\}
\newcommand{\gridResolutionTableBody}{%
Camel & 27.9 & 0.234 & 28.4 & 0.168 & 28.7 & 0.132 & 38.0 & 0.105 & 90.3 & 0.104 \\
Dragon & 51.7 & 0.280 & 51.3 & 0.186 & 52.8 & 0.147 & 61.4 & 0.109 & 112 & 0.098 \\
Kitten & 186 & 0.323 & 185 & 0.271 & 189 & 0.237 & 199 & 0.210 & 262 & 0.202 \\
Lucy & 401 & 0.422 & 401 & 0.340 & 403 & 0.298 & 427 & 0.245 & 484 & 0.249 \\}

\newcommand{\appIcpTableBody}{%
DP-NNS~\cite{dpnns} & \underline{9.44} & \underline{7.07} & \underline{4.94} & \underline{5.73} \\
nanoflann & 17.1 & 13.8 & 9.31 & 12.1 \\
R$^*$-tree & 37.4 & 26.5 & 18.2 & 23.0 \\
BD-tree & 45.2 & 41.0 & 24.3 & 34.9 \\
Octree & 101 & 75.6 & 56.1 & 67.2 \\
PCL $k$-d tree & 121 & 72.2 & 50.6 & 79.2 \\
Ours & \textbf{\underline{5.55}} & \textbf{\underline{4.04}} & \textbf{\underline{2.87}} & \textbf{\underline{3.27}} \\}

\begin{document}
\bstctlcite{IEEEexample:BSTcontrol}

\title{Delaunay-Based Nearest Neighbor Search\texorpdfstring{\\}{} with Grid-Accelerated Queries\texorpdfstring{\\}{} and Parallel Construction}

\author{Pengfei~Wang,
        Shiqing~Xin,
        Changhe~Tu
}

\markboth{}%
{Wang \MakeLowercase{\textit{et al.}}: Delaunay-Based Nearest Neighbor Search with Grid-Accelerated Queries and Parallel Construction}

\IEEEtitleabstractindextext{%
\begin{abstract}
Nearest-neighbor search is a fundamental primitive in computer graphics, with applications in distance field computation, surface reconstruction, and point cloud registration. In many graphics applications, data points lie on a surface manifold embedded in 3D space, whereas queries are distributed throughout the surrounding volume—a setting in which spatial partitioning structures such as $k$-d trees and grids can become inefficient. Dynamic-programming-based nearest-neighbor search (DP-NNS) offers an alternative: it traverses a successor DAG that records the history of incremental Delaunay construction and often outperforms spatial partitioning structures. However, its sequential construction is slow, and naive parallelization degrades query performance, limiting its practical adoption. We accelerate both queries and construction in DP-NNS. For queries, we introduce a grid-based initialization strategy that precomputes valid starting points on the successor DAG, shortening search paths while preserving exactness and enabling a twofold speedup. For construction, we analyze how parallel insertion affects query cost and develop a parallel algorithm that preserves logarithmic expected query cost under stated geometric assumptions. Our parallel construction algorithm achieves a 10$\times$ speedup over the sequential baseline while maintaining nearly the same query cost. We further reorder query points to improve cache locality. Extensive experiments demonstrate that our method outperforms widely used libraries in nearest-neighbor query performance.

\end{abstract}

}

\maketitle
\IEEEdisplaynontitleabstractindextext
\IEEEpeerreviewmaketitle

\begin{figure}[htbp]
    \centering
    \includegraphics[width=0.97\linewidth]{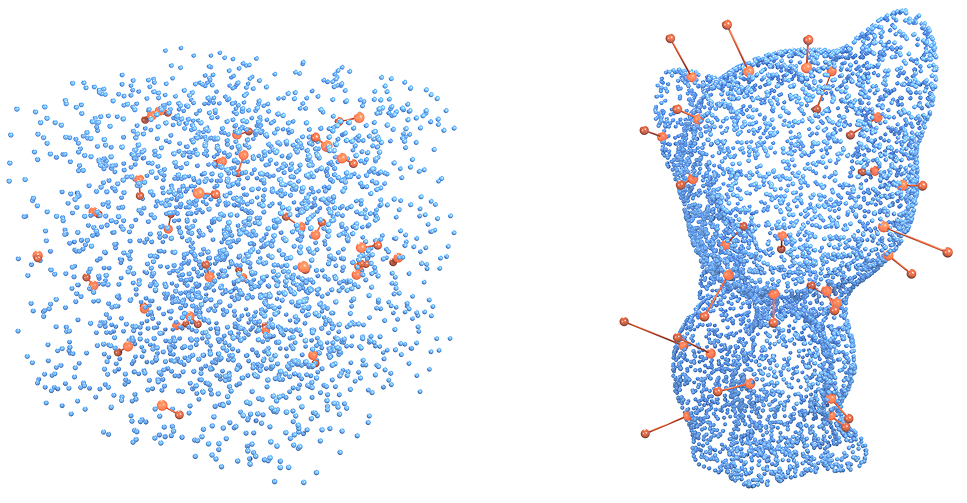}
\caption{Nearest-neighbor queries for two data distributions. Left: data points distributed uniformly throughout a 3D volume. Right: data points lying on a manifold embedded in 3D space, with queries distributed throughout the surrounding volume.}
    \label{fig:nn_example}
\end{figure}

\IEEEraisesectionheading{\section{Introduction}\label{sec:introduction}}
\IEEEPARstart{N}{earest-neighbor} search is a fundamental operation in computer graphics and geometry processing.
In many applications, data points are sampled from a surface, whereas query points are distributed throughout the surrounding volume, as illustrated in Figure~\ref{fig:nn_example} (right).
For example, implicit surface reconstruction evaluates distances to surface samples at grid vertices or other spatial locations~\cite{hoppe1992surface,Wang2021NeuralIMLSSI}; neural implicit representations use volumetric queries during training~\cite{ma2021neuralpull}; and point-based rendering searches for nearby surface points at locations along viewing rays~\cite{Ost2021NeuralPL,Chang2023PointersectNR,Chen2023NeuralEditorEN}. These applications often require millions of nearest-neighbor queries, making search efficiency critical to their overall performance.

Most nearest-neighbor indices, including $k$-d trees~\cite{bentley1975}, R-trees~\cite{10.1145/602259.602266,rstarTREE}, and bounding volume hierarchies~\cite{arborx2}, organize space hierarchically and prune branches during traversal. However, when data points lie on a surface manifold embedded in 3D space while queries fill the surrounding volume, bounding regions enclose substantial empty space; for distant queries, this reduces pruning efficiency and slows down or even degrades query performance. Uniform grids~\cite{bentley1980optimal} provide direct access to nearby cells and work well for uniformly distributed points. For manifold-distributed data, however, most grid cells remain empty, forcing distant queries to traverse a massive number of empty cells and causing severe performance degradation. DP-NNS~\cite{dpnns} instead exploits the history of incremental Delaunay construction: each point records later insertions that become its Delaunay neighbors. These successor lists form a directed acyclic graph (the successor DAG), along which queries advance directly toward their nearest neighbors. For volume queries against surface manifolds, DP-NNS often achieves several-fold speedups over tree-based indices~\cite{dpnns}.

DP-NNS, however, achieves these query-time gains at a high construction cost. Sequential incremental Delaunay construction is substantially slower than building a spatial tree, so DP-NNS becomes worthwhile only for sufficiently large query batches. Existing parallel Delaunay schemes, such as CGAL's parallel insertion~\cite{cgal:hs-chdt3-24a,batista2010}, accelerate construction but can degrade query performance by altering the insertion order. This order determines the successor DAG and strongly affects its query cost: in the worst case, inserting points in decreasing order of distance to a query forces the walk to visit every point.

We improve the overall efficiency of DP-NNS by accelerating both queries and construction. For queries, each search follows a path through the successor DAG to the nearest neighbor. We observe that starting from any point on this path preserves the result. We therefore identify shared intermediate points for queries within spatial regions and precompute them as new starting points, shortening search paths and accelerating queries. For construction, the sequential process is slow, while conventional parallel insertion can increase query cost by disrupting the coarse-to-fine insertion order. We analyze this dependence and show that, under our geometric assumptions, preserving the order across randomly sampled levels~\cite{amenta2003brio} retains logarithmic expected query cost, regardless of the order within each level. We thus process levels sequentially while inserting points within each level in parallel~\cite{marot2019onemachine}, accelerating construction while maintaining query efficiency. We further improve cache locality by sorting query batches along a Morton curve. While the query initialization targets 1-NN search, parallel construction and query reordering also benefit Manifold $k$-NN~\cite{manifoldknn}.

\begin{figure}[t]
    \centering
    \includegraphics[width=\columnwidth]{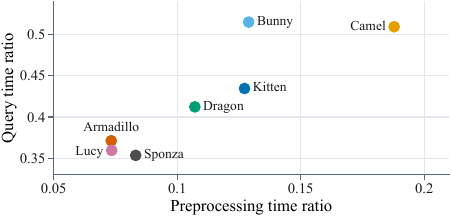}
\caption{Preprocessing and average 1-NN query times relative to the original DP-NNS~\cite{dpnns} on several classic benchmark models (lower is better). Each ratio is our time divided by the corresponding DP-NNS time. Our method is significantly faster than the original DP-NNS in both preprocessing and query execution.}
    \label{fig:time_ratio}
\end{figure}

Figure~\ref{fig:time_ratio} demonstrates the effectiveness of our query initialization and parallel construction. Our main contributions are:
\begin{itemize}
\item A grid-based initialization strategy that accelerates nearest-neighbor queries while preserving exactness.
\item A theoretically supported parallel construction algorithm that preserves logarithmic expected query cost under our geometric assumptions.
\item An efficient implementation with query reordering that outperforms widely used libraries for volumetric nearest-neighbor queries on surface data.
\end{itemize}

\section{Related Work}
\label{sec:related_work}

\subsection{Nearest Neighbor Search}
\label{sec:related_knn}

\paragraph{Space-partitioning indices}
Most exact nearest-neighbor indices partition data or space into regions and use distance lower bounds to prune regions that cannot contain a closer point. The $k$-d tree~\cite{bentley1975,friedman1977} recursively partitions the point set with axis-aligned planes. A query descends to a leaf and then backtracks through nodes whose bounding regions intersect the current search ball. Ball trees and vantage-point trees~\cite{omohundro1989,yianilos1993} instead organize points using metric balls or distance-based partitions. Uniform grids and spatial hashing provide direct access to nearby cells and are effective for uniformly distributed points, whereas octrees~\cite{meagher1982} adaptively subdivide occupied space. R-trees and R$^*$-trees~\cite{10.1145/602259.602266,rstarTREE} group points within potentially overlapping bounding boxes, while bounding volume hierarchies (BVHs)~\cite{arborx2} organize points and other geometric primitives in nested bounding regions.

The performance of these structures hinges on how tightly their bounding volumes fit the underlying geometry. When data points sample a surface while queries fill the ambient volume, spatial cells and bounding boxes enclose vast empty space and overlap around thin or curved regions. For queries far from the surface, the initial nearest-neighbor distance bound is large, rendering branch-and-bound pruning largely ineffective; as a result, tree traversals often inspect a substantial fraction of the hierarchy. Parallel execution reduces wall-clock runtime but does not eliminate this redundant work.

\paragraph{Graph-based search}
Graph-based methods do not prune regions but move from point to point toward the query. Walking in a triangulation~\cite{devillers2002walking} and greedy routing~\cite{bose2004online} follow Delaunay edges toward a target. The Delaunay tree~\cite{boissonnat1993delaunaytree} and the history graph of randomized incremental construction~\cite{guibas1992randomized} locate a point by descending through simplices that later insertions destroyed, while the Delaunay hierarchy~\cite{devillers2002hierarchy} walks through triangulations of nested random samples. For approximate search in high dimensions, HNSW~\cite{malkov2020hnsw} searches navigable graphs whose random layers shrink geometrically, and SONG~\cite{zhao2020song}, GGNN~\cite{groh2023ggnn}, and CAGRA~\cite{ootomo2024cagra} traverse proximity graphs on the GPU using one warp or thread block per query.

DP-NNS~\cite{dpnns} exploits the history of incremental Delaunay triangulation for nearest-neighbor search. Each newly inserted point is appended to the successor lists of its Delaunay neighbors, recording candidates that may replace an earlier point as the nearest neighbor. A query traverses these lists toward progressively closer points rather than backtracking through a tree, making the method particularly effective when data points lie on a manifold and queries occupy the surrounding volume. 

\subsection{Delaunay Construction}
\label{sec:related_parallel}

Delaunay triangulations are commonly constructed by incremental insertion~\cite{bowyer1981,watson1981}. Each new point is inserted by identifying the tetrahedra that violate the Delaunay condition and retriangulating the affected region to restore it. 
Construction efficiency depends strongly on the insertion order. Random insertion provides theoretical guarantees on the expected time complexity but often has poor memory locality, whereas ordering points along a space-filling curve improves locality without the same complexity guarantees. Biased randomized insertion order (BRIO)~\cite{amenta2003brio} combines these benefits by organizing points into random groups of geometrically increasing size and ordering each group along a space-filling curve. This approach improves spatial locality while retaining the theoretical efficiency of randomized incremental construction.

Parallel methods accelerate Delaunay construction by inserting points concurrently while coordinating conflicts between local updates. Shared-memory approaches use locking or spatial partitioning to manage concurrent insertions~\cite{cgal:hs-chdt3-24a,batista2010,kohout2005parallel,blandford2006compact,foteinos2011dynamic,marot2019onemachine}, while GPU methods exploit large-scale parallelism~\cite{cao2014gdel3d,Gao2025Parallel3D,Gao2026LocalDB}. Other approaches preserve a sequential random insertion order through parallel scheduling~\cite{blelloch2012,blelloch2016}. These methods primarily target the final triangulation, which is independent of insertion order under general position. The insertion history, however, depends on this order and determines the successor DAG used by DP-NNS. Section~\ref{sec:theory} analyzes how this dependence affects query cost.

\section{Preliminaries}
\label{sec:prelim}

Our method builds on DP-NNS~\cite{dpnns}. We first review how incremental Delaunay construction defines a successor DAG and how nearest-neighbor queries traverse this graph.

\paragraph{Notation}
Let $\mathcal{P}\subset\mathbb{R}^3$ contain $n$ points, indexed as $p_1,\ldots,p_n$ in their Delaunay insertion order. The rank of $p_i$ is $r(p_i)=i$. For each point $p$, define
\begin{equation}
   \mathcal{P}_{<p}=\{x\in\mathcal{P}: r(x)<r(p)\},
\qquad
\mathcal{P}_{\le p}=\mathcal{P}_{<p}\cup\{p\}. 
\end{equation}
We denote the Delaunay triangulation of a point set $X$ by $\mathrm{DT}(X)$ and the Voronoi cell of $x\in X$ by $\mathrm{Cell}(x;X)$. The notation $xy\in\mathrm{DT}(X)$ indicates that $x$ and $y$ share a Delaunay edge, or equivalently, that their Voronoi cells share a face. We assume general position to avoid degeneracies. In practice, CGAL handles degenerate inputs through symbolic perturbation~\cite{devillers2011perturbations}.

\begin{figure}[t]
    \centering
    \includegraphics[width=0.99\linewidth]{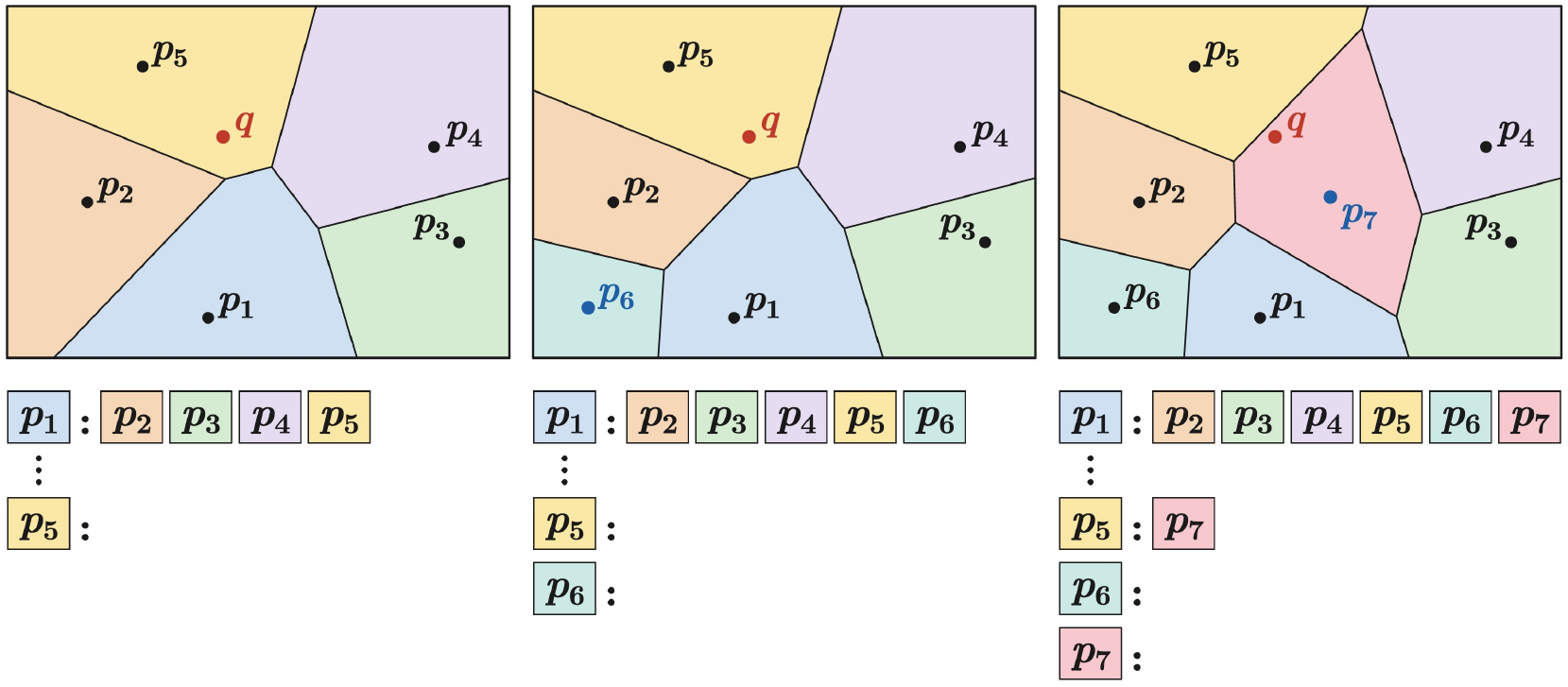}
    \caption{Successor lists on a two-dimensional example. Left: the Voronoi diagram of $p_1,\ldots,p_5$; the query $q$ lies in the cell of $p_5$. Middle: inserting $p_6$ does not change the cell of $p_5$, so $\Phi_6(q)=\Phi_5(q)=p_5$. Right: inserting $p_7$ shrinks the cell of $p_5$ and takes $q$, so $\Phi_7(q)=p_7$. Below each diagram are the successor lists at that time: each point is appended to the lists of its Delaunay neighbors when it is inserted.}
    \label{fig:succ_lists}
\end{figure}

\paragraph{Observation}
For a query point $q$, let $\Phi_m(q)$ be the nearest neighbor of $q$ in $\{p_1,\ldots,p_m\}$. It follows the recursion
\begin{equation}
\Phi_m(q)=\begin{cases}
p_m, & m=1 \text{ or } \|q-p_m\|<\|q-\Phi_{m-1}(q)\|,\\
\Phi_{m-1}(q), & \text{otherwise}.
\end{cases}
\label{eq:prefix_nn}
\end{equation}
Thus, $\Phi_n(q)$ is the nearest neighbor in $\mathcal{P}$. Let $c=\Phi_{m-1}(q)$ be the nearest neighbor before $p_m$ is inserted; by definition, $q\in\mathrm{Cell}(c;\mathcal{P}_{<p_m})$. Inserting $p_m$ creates a new cell by taking regions from the cells of its Delaunay neighbors. Therefore, $p_m$ can replace $c$ as the nearest neighbor only if the two points become Delaunay neighbors~\cite{dpnns}:
\begin{equation}
c\,p_m\notin\mathrm{DT}(\mathcal{P}_{\le p_m}) \;\Longrightarrow\; \Phi_m(q)=\Phi_{m-1}(q)=c.
\label{eq:observation}
\end{equation}
Figure~\ref{fig:succ_lists} illustrates this in two dimensions.

\paragraph{Successor DAG}
This observation allows DP-NNS to restrict the possible nearest-neighbor updates. For each point $x$, it records the later points that become Delaunay neighbors of $x$ upon insertion:
\begin{equation}
\mathrm{succ}(x)=\{p\in\mathcal{P}:\ r(p)>r(x),\ xp\in\mathrm{DT}(\mathcal{P}_{\le p})\},
\label{eq:succ}
\end{equation}
These successors are stored in increasing rank order. During construction, each new point is appended to the successor lists of its existing Delaunay neighbors. The resulting directed edges form the \emph{successor DAG}, which is acyclic because every edge leads to a point of higher rank.

\paragraph{Nearest-neighbor search}
Starting at $c_0=p_1$, a query scans the successors of its current point in rank order and moves to the first one closer to $q$:
\begin{equation}
c_{t+1}=
\mathop{\arg\min}_{\substack{p\in\mathrm{succ}(c_t)\
|q-p|<|q-c_t|}}
r(p).
\label{eq:dpquick}
\end{equation}
The walk terminates when no successor is closer. By Eq.~\eqref{eq:prefix_nn} and~\eqref{eq:observation}, each move identifies the earliest subsequent insertion that changes the nearest neighbor. The walk therefore visits exactly the \emph{transition sites}
\begin{equation*}
\mathcal{T}(q)=\{p_m\in\mathcal{P}:\ \Phi_m(q)=p_m\},
\end{equation*}
in increasing rank order and ends at $\Phi_n(q)$.

\begin{figure}[t]
    \centering
    \includegraphics[width=0.99\linewidth]{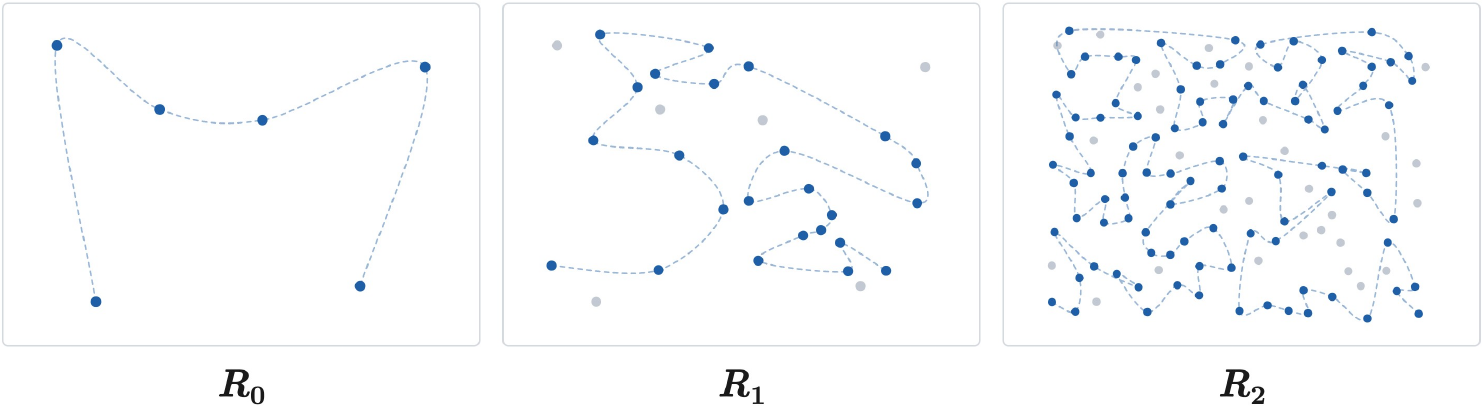}
\caption{Levels of a biased randomized insertion order on a two-dimensional example. The shuffled points are split into levels $R_0$, $R_1$ and $R_2$, inserted one after another; within each level, the points are inserted along a Hilbert curve (dashed). Grey points belong to earlier levels, so that $S_1=R_0$ and $S_2=R_0\cup R_1$. For readability, each level here has four times as many points as all earlier levels, instead of about seven in CGAL.}
    \label{fig:levels2d}
\end{figure}

\paragraph{Multilevel insertion order}
DP-NNS inserts points in an order $\sigma$ generated by CGAL's spatial sort~\cite{}, which follows a biased randomized insertion order (BRIO)~\cite{amenta2003brio}. This ordering combines random sampling with spatial locality. The point set $\mathcal{P}$ is first randomly shuffled and partitioned into $O(\log n)$ levels $R_0,\ldots,R_L$ of geometrically increasing size. Each prefix $S_l=R_0\cup\cdots\cup R_{l-1}$ therefore forms a uniform random sample of $\mathcal{P}$. Within each level, points are sorted along a Hilbert curve, and the levels are concatenated to form $\sigma$, as illustrated in Fig.~\ref{fig:levels2d}. This structure retains the asymptotic guarantees of randomized incremental construction while improving spatial locality~\cite{amenta2003brio}. Its effect on the successor DAG's query cost motivates our analysis in Section~\ref{sec:theory}.

\section{METHOD}
Given a point set $\mathcal{P}$ lying on a surface manifold in 3D space, we accelerate both queries and successor-DAG construction in DP-NNS~\cite{dpnns}. For queries, we use a uniform grid to precompute starting points shared by all query paths within each cell, allowing queries to skip early traversal steps while preserving exactness. For construction, we show that, under our geometric assumptions, preserving the insertion order across randomly sampled levels suffices to retain $O(\log n)$ expected query cost. This result allows insertions within each level to proceed in parallel without sacrificing query efficiency.

\begin{figure}[t]
    \centering
    \includegraphics[width=\columnwidth]{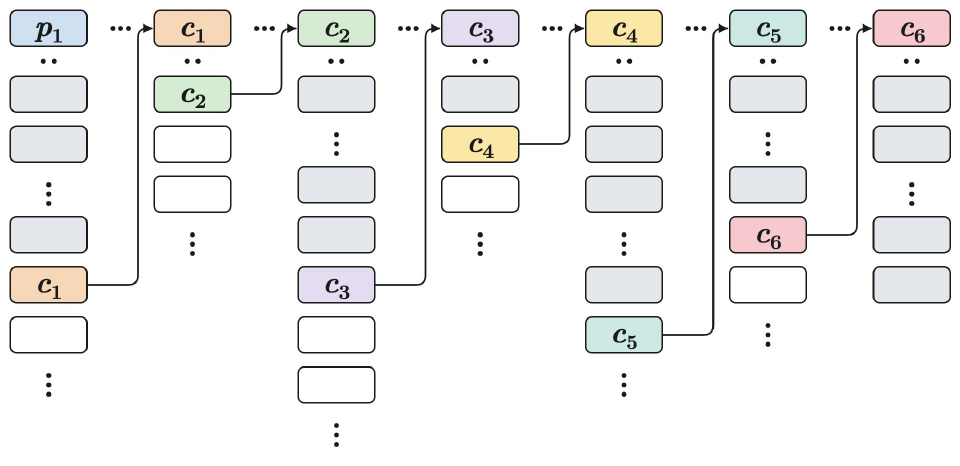}
        \label{fig:querypath}
 \caption{A search path $p_1,c_1,\dots,c_6$ through the successor DAG, with successor lists shown below each point. Starting from any point on the path preserves the final nearest neighbor while skipping earlier search steps.}

\end{figure}

\subsection{Grid-Accelerated Search Initialization}
\label{sec:Grid-Accelerate}
The recurrence in Eq.~\ref{eq:prefix_nn} allows a nearest-neighbor query to resume from any known intermediate state $\Phi_m(q)$, since subsequent updates depend only on the current state and the remaining points. Thus, knowing $\Phi_m(q)$ in advance allows the query to skip all preceding updates.

To apply this observation to the successor-DAG traversal in Eq.~\ref{eq:dpquick}, we define the \emph{query path} as the set of points that become the nearest neighbor upon insertion, as illustrated in Fig.~\ref{fig:querypath}:
\[
\mathcal{T}(q) = \{p_m \in \mathcal{P} : \Phi_m(q) = p_m\}.
\]
The traversal visits these points in insertion order. Any point $p_m \in \mathcal{T}(q)$ therefore provides a valid starting point: initializing the search at $p_m$ preserves the remaining traversal and its final result while bypassing the earlier steps. To accelerate queries within a spatial region, we seek a starting point shared by all their query paths. For a convex polytope, such a point can be identified by examining only the paths of its vertices.

\begin{lemma}
Let $\Omega \subset \mathbb{R}^3$ be a convex polytope with vertices $v_1, \dots, v_m$.
Then
\[
\bigcap_{i=1}^{m} \mathcal{T}(v_i)
\subseteq \mathcal{T}(q)
\qquad \text{for every } q \in \Omega.
\]
\label{lemma:polytope_walk}
\end{lemma}

\begin{figure}[t]
    \centering
    \includegraphics[width=\columnwidth]{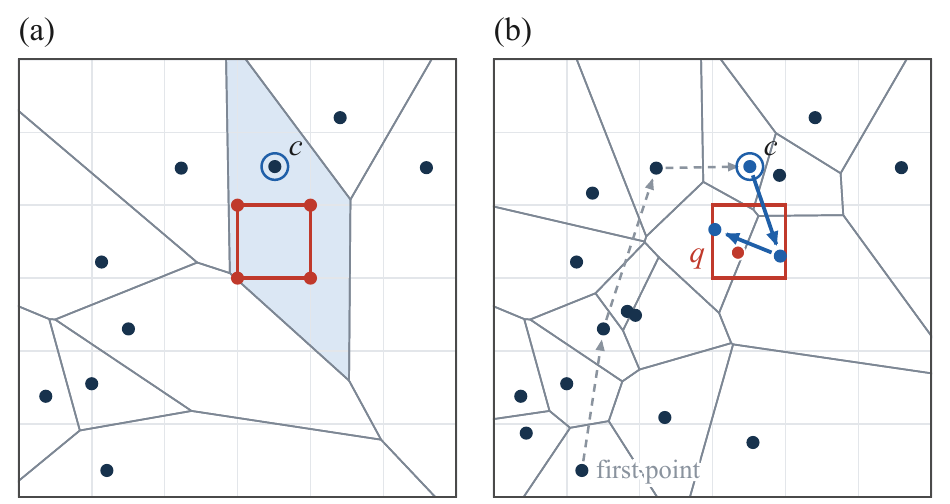}
    \caption{Grid-based query initialization in 2D. (a) At the insertion of $c$, all vertices of the marked grid cell lie in its Voronoi cell (shaded); by convexity, the entire grid cell lies inside as well. (b) Any query $q$ within the grid cell can start directly from $c$. Blue arrows show the grid-accelerated query path, while gray dashed arrows indicate the skipped prefix of the original path.}
    \label{fig:grid}
\end{figure}

\begin{proof}
Let $c \in \bigcap_{i=1}^{m} \mathcal{T}(v_i)$.
By the definition of $\mathcal{T}$, for any query point $x$,
\[
c \in \mathcal{T}(x)
\iff \Phi_{r(c)}(x) = c
\iff x \in \mathrm{Cell}(c; \mathcal{P}_{\le c}).
\]
Thus, every vertex $v_i$ belongs to $\mathrm{Cell}(c; \mathcal{P}_{\le c})$.
By the convexity of Voronoi cells (as shown in Fig.~\ref{fig:grid}a),
\[
\Omega = \operatorname{conv}\{v_1,\dots,v_m\}
\subseteq \mathrm{Cell}(c; \mathcal{P}_{\le c}).
\]
Thus, every $q \in \Omega$ lies in $\mathrm{Cell}(c; \mathcal{P}_{\le c})$, so $c$ is its nearest neighbor in $\mathcal{P}_{\le c}$. Hence, $\Phi_{r(c)}(q) = c$, which implies $c \in \mathcal{T}(q)$ and proves the inclusion.
\end{proof}

Lemma~\ref{lemma:polytope_walk} allows us to partition the query domain into convex cells and precompute a valid starting point shared by all queries within each cell.
At query time, we first identify the cell containing the query point and then initialize the DP-NNS traversal at the starting point stored in that cell (see Fig.~\ref{fig:grid}).
Since cell location contributes to the total query cost, it must be performed efficiently.
We therefore use a uniform grid to partition the domain into $G^3$ cells, enabling direct access to the corresponding starting point through $O(1)$ cell lookup.

To compute these starting points, we perform a nearest neighbor query at each of the $G^3$ grid vertices $v$ and record its query path $\mathcal{T}(v)$. For a cell $C$ with vertices $v_1,\dots,v_8$, Lemma~\ref{lemma:polytope_walk} ensures that every point shared by these eight paths is a valid starting point for all queries in $C$. Since points along each path are ordered by rank, we select the shared point of highest rank to bypass the largest number of preceding steps:
\[
s(C) = \arg\max_{c \in \bigcap_{i=1}^{8} \mathcal{T}(v_i)} r(c).
\]
This intersection is nonempty because $p_1$ belongs to every query path. The selected point $s(C)$ is stored in cell $C$ and serves as the starting point for subsequent queries within that cell.

\subsection{Level-Consistent Parallel Construction}
\label{sec:theory}
While the final Delaunay triangulation depends solely on point locations and is invariant to the insertion order, the successor DAG used in DP-NNS is not; it relies heavily on the insertion sequence during incremental construction, which dictates query efficiency.
Conventional parallelization disrupts this sequential order and leads to severe query performance degradation.
In this work, we prove that under a constrained parallelization strategy, we can accelerate construction while preserving the expected logarithmic query cost—ensuring that parallelization does not compromise query speed.

\begin{figure}[t]
    \centering
    \includegraphics[width=\columnwidth]{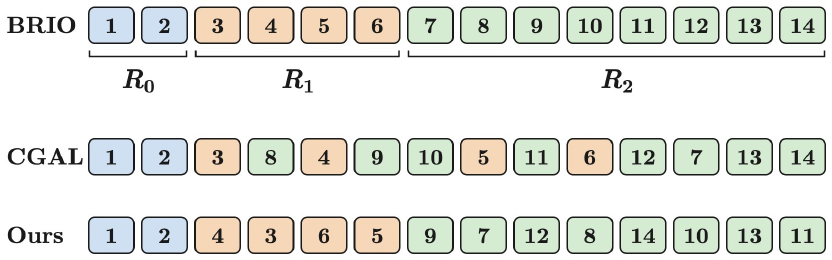}
\caption{Comparison of insertion schedules: initial BRIO order (top), conventional parallel insertion in CGAL (middle), and our level-synchronous parallel construction (bottom). Colors denote levels ($R_0, R_1, R_2$). While CGAL interleaves points across levels, our method preserves a level-consistent order while enabling parallel execution within each level.}
    \label{fig:order}
\end{figure}

We build upon the multilevel BRIO insertion schedule described in Section~3, which combines random sampling with spatial locality.
Recall that the point set $\mathcal{P}$ is first randomly shuffled and partitioned into $L+1 = O(\log n)$ levels $R_0, \dots, R_L$ of geometrically increasing size.
For each level $l \ge 1$, let
\[
S_l = R_0 \cup \cdots \cup R_{l-1}
\]
denote the points in all preceding levels.
Because of the initial random shuffle, each $S_l$ forms a uniform random sample of $\mathcal{P}$, and the level sizes satisfy
\[
|R_l| \le \gamma |S_l|,
\]
where $\gamma \ge 1$ is a constant controlling the growth rate (see Fig.~\ref{fig:order}, top).

We call an insertion order \emph{level-consistent} if all points in $R_{l-1}$ are inserted before any point in $R_l$ for every $l \ge 1$, while allowing arbitrary insertion orders within each level, as illustrated in Fig.~\ref{fig:order} (Ours).
We further assume that the data points in $\mathcal{P}$ follow a homogeneous Poisson point process.
Under this assumption, each point has a constant expected number of Delaunay neighbors, denoted by $\delta$, which is also consistent with empirical observations~\cite{mei2018degreedistributiondelaunaytriangulations}.
These conditions yield the following theorem.

\begin{theorem}
\label{thm:level_consistent}
Under any level-consistent insertion order, a nearest-neighbor query for $q$ evaluates distances to $O(1)$ points per level in expectation, and $O(\log n)$ points in total.
\end{theorem}

\begin{proof}
A nearest-neighbor query evaluates the distance to a point $p \in R_l$ ($p \ne p_1$) if and only if $p$ is a successor of the current nearest neighbor $c = \Phi_{r(p)-1}(q)$, i.e., $cp \in \mathrm{DT}(\mathcal{P}_{\le p})$.
Under a level-consistent order, $S_l \subseteq \mathcal{P}_{<p} \subseteq S_l \cup R_l$.
Because removing points never destroys existing Delaunay edges among the remaining points, $cp \in \mathrm{DT}(\mathcal{P}_{\le p})$ implies $cp \in \mathrm{DT}(S_l \cup \{c, p\})$.
Depending on whether $c$ lies in $S_l$ or $R_l$, we bound the expected number of distance evaluations in $R_l$ in two disjoint cases:

\noindent\textbf{Case 1 ($c \in S_l$):}
Since all points in $S_l$ are inserted before $R_l$, $c$ is the nearest neighbor of $q$ in $S_l$.
Evaluating the distance to $p \in R_l$ requires $cp \in \mathrm{DT}(S_l \cup \{p\})$.
Because the points are randomly shuffled, $c$ has $\delta$ Delaunay neighbors in expectation and $p$ is a random point, so $p$ is adjacent to $c$ in $\mathrm{DT}(S_l \cup \{p\})$ with probability $O(\delta / |S_l|)$.
Multiplying by $|R_l|$ yields an expected count of
\begin{equation}
|R_l| \cdot O\!\left(\frac{\delta}{|S_l|}\right) \le \gamma |S_l| \cdot O\!\left(\frac{\delta}{|S_l|}\right) = O(\gamma \delta) = O(1).
\end{equation}

\noindent\textbf{Case 2 ($c \in R_l$):}
Since all points in $S_l$ are inserted before $c$, $c$ is closer to $q$ than all points in $S_l$.
We bound the expected number of pairs $(c, p)$ satisfying $cp \in \mathrm{DT}(S_l \cup \{c, p\})$ in two steps.
First, by the random shuffle, a point in $R_l$ is closer to $q$ than all points in $S_l$ with probability $1 / (|S_l| + 1)$, so $R_l$ contains at most $\gamma |S_l| / (|S_l| + 1) \le \gamma$ such points $c$ in expectation.
Second, each such $c$ is the nearest neighbor of $q$ in the random sample $S_l \cup \{c\}$ and thus has $\delta$ Delaunay neighbors in expectation.
As in Case~1, a random point $p \in R_l$ is adjacent to $c$ in $\mathrm{DT}(S_l \cup \{c, p\})$ with probability $O(\delta / |S_l|)$, accounting for at most $|R_l| \cdot O(\delta / |S_l|) = O(\gamma \delta)$ points $p$ in expectation.
Multiplying by the expected number of candidates $c$ yields
\begin{equation}
\gamma \cdot O(\gamma \delta) = O(\gamma^2 \delta) = O(1).
\end{equation}

Combining both cases, the expected number of distance evaluations in $R_l$ is bounded by $O(1) + O(1) = O(1)$.
Summing over all $L + 1 = O(\log n)$ levels $R_0, \dots, R_L$ yields an $O(\log n)$ expected total cost.
\end{proof}

This analysis relies on two properties: the randomness introduced by the shuffle before the points are split into levels, and the level-consistent insertion order. If either is violated, the complexity bound no longer holds. These two properties form the theoretical basis of the algorithm designed below.

\subsection{Implementation of Parallel DAG Construction}
\label{sec:ParallelDAGConstruction}

Guided by Theorem~\ref{thm:level_consistent}, our parallel construction processes the levels $R_0, \dots, R_L$ sequentially while parallelizing insertions within each level. 
We first compute the BRIO sequence $\sigma$ via a parallel spatial sort. 
Small initial levels are inserted sequentially as a prefix, whereas for remaining levels, threads dynamically fetch contiguous blocks along $\sigma$ using a shared atomic counter to balance workload while preserving spatial locality. 
Enforcing a synchronization barrier between levels guarantees a strictly level-consistent insertion order.

Within each level, concurrent threads insert different blocks simultaneously. 
To ensure a valid sequential equivalent, conflicting insertions operating on overlapping mesh regions must be coordinated. 
Unlike CGAL, which relies on auxiliary grid locking~\cite{batista2010}, we attach locks directly to the triangulation vertices: inserting a point locks all vertices of its visited and conflicting tetrahedra. 

To prevent deadlocks, threads acquire locks in priority order, waiting only for locks held by lower-priority threads. 
If a higher-priority lock blocks progress, the thread releases all acquired locks, defers the current point, and resumes with subsequent points in its block; deferred points are retried after the block finishes. 
Once locks are secured, the thread updates the triangulation, assigns a rank from a shared counter, and records its successor edges, which are assembled into the final successor lists after all levels complete.

\subsection{Locality-Aware Parallel Queries}
\label{sec:ParallelQueries}
Each query independently accesses the uniform grid to determine its starting point and traverses the successor DAG. The processing order of queries affects memory access locality: spatially nearby queries often visit common vertices and examine overlapping successor lists. We exploit this locality by sorting each query batch along a 3D Morton curve using radix sort. This ordering groups nearby queries together, increasing the likelihood that vertices and successor lists accessed by one query remain in the cache for subsequent queries, thereby reducing repeated memory accesses during traversal.

\begin{table}[t]
  \centering
  \caption{The seven test models and their numbers of points.}
  \label{fig:seven_models}
  \setlength{\tabcolsep}{2.2pt}
  \begin{tabular}{lccccccc}
    \toprule
            & Camel & Bunny & Dragon & Kitten & Armadillo & Lucy & Sponza \\
    \midrule
    Points  & 29K & 73K & 100K & 291K & 726K & 1.02M & 1.31M \\
    \bottomrule
  \end{tabular}
\end{table}

\section{Evaluation}
\label{sec:evaluation}

\begin{table}[t]
    \centering
    \caption{Average nearest-neighbor query time ($\mu$s). Best and second-best results are \textbf{\underline{bold underlined}} and \underline{underlined}. Both DP-NNS$^1$ and DP-NNS$^2$ use the original DP-NNS query, evaluated on the DAGs built by sequential insertion and by CGAL's parallel insertion, respectively.}
    \label{tab:queries}
    \setlength{\tabcolsep}{2.2pt}
    \scriptsize
    \begin{tabular}{l rrrrrrr}
        \toprule
        Method & Camel & Bunny & Dragon & Kitten & Armadillo & Lucy & Sponza \\
        \midrule
\queryCpuOneTableBody
        \bottomrule
    \end{tabular}
\end{table}

\begin{table}[t]
    \centering
    \caption{Average distance evaluations per query on DAGs built via sequential insertion, CGAL's parallel insertion, and our level-consistent parallel construction, starting from either $p_1$ (standard query) or the grid cell's representative $s(C)$ (grid-accelerated query). }
    
    \label{tab:query_cost}
    \setlength{\tabcolsep}{3pt}
    \footnotesize
    \begin{tabular}{l cc cc cc}
        \toprule
        & \multicolumn{2}{c}{Sequential} & \multicolumn{2}{c}{Conventional} & \multicolumn{2}{c}{Level-consistent} \\
        \cmidrule(lr){2-3}\cmidrule(lr){4-5}\cmidrule(lr){6-7}
        Model & $p_1$ & $s(C)$ & $p_1$ & $s(C)$ & $p_1$ & $s(C)$ \\
        \midrule
\queryCostTableBody
        \bottomrule
    \end{tabular}
\end{table}
\subsection{Experimental Setup}
\label{sec:exp_setup}

\paragraph{Platform and data}
All experiments run on a workstation with an Intel Core i7-14700K, which has 8 performance and 12 efficiency cores, 64~GB of memory and an NVIDIA GeForce RTX~4090 with 24~GB of memory, under Windows~11. Our implementation uses CGAL~5.6.1~\cite{cgal:hs-chdt3-24a} with exact predicates for the Delaunay insertion and oneTBB for threading. We evaluate our method on seven classic 3D model point clouds (Table~\ref{fig:seven_models}) as well as three large-scale datasets: the Thingi10K dataset~\cite{thingi10k}, the ABC dataset~\cite{Koch_2019_CVPR}, and the MedShapeNet dataset~\cite{medshapenet}.

\paragraph{Queries and timing}
Unless stated otherwise, we follow DP-NNS~\cite{dpnns} and generate 100K queries for each model uniformly within its bounding box, expanded by half of its side length along each direction (yielding a query domain with eight times the original bounding-box volume).
For multi-threaded construction, we use 20 threads for all methods.
All reported query times for our method include the overhead of sorting the query points.

\paragraph{Compared methods}
We compare against the original implementation of DP-NNS~\cite{dpnns}, the nanoflann $kd$-tree~\cite{nanoflann}, the Boost R$^*$-tree~\cite{boostgeometry}, the PCL octree~\cite{pcl}, the ANN BD-tree~\cite{annlibrary}, SciPy's \texttt{cKDTree}~\cite{virtanen2020scipy}, and the ArborX BVH~\cite{arborx2}.
All methods are configured for exact search, with the remaining parameters set to their defaults.

\begin{figure}[t]
    \centering
    \includegraphics[width=\columnwidth]{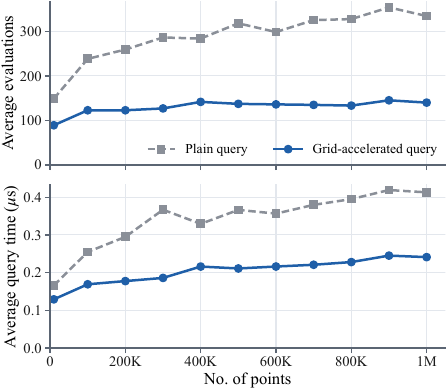}
    \caption{Comparison of standard (from $p_1$) and grid-accelerated queries on the Lucy model across varying densities using our level-consistent parallel construction: average distance evaluations (top) and query time (bottom).}
    \label{fig:grid_scaling}
\end{figure}

\begin{table}[t]
    \centering
    \caption{Construction time (ms). Methods above the dashed line are sequential-only, whereas those below are parallelized. DP-NNS$^1$: original sequential construction; DP-NNS$^2$: DAG construction via CGAL's parallel insertion.}
    \label{tab:construction}
    \setlength{\tabcolsep}{2.2pt}
    \scriptsize
    \begin{tabular}{l rrrrrrr}
        \toprule
        Method & Camel & Bunny & Dragon & Kitten & Armadillo & Lucy & Sponza \\
        \midrule
\constructionAllTableBody
        \bottomrule
    \end{tabular}
\end{table}
\subsection{Query Performance}
\label{sec:exp_query}

Table~\ref{tab:queries} compares nearest-neighbor query times across the seven benchmark models, where our method achieves an approximate $2\times$ speedup over the fastest baseline. To isolate the effect of grid-accelerated search initialization, Table~\ref{tab:query_cost} reports the average number of distance evaluations per query with and without initialization across different construction strategies. Across all seven models, this initialization substantially reduces distance evaluations compared to the standard traversal from $p_1$, confirming its effectiveness in shortening search paths. Fig.~\ref{fig:grid_scaling} further evaluates this behavior on the Lucy model across varying densities, demonstrating consistent reductions in both distance evaluations and query time.

\begin{figure}[t]
    \centering
    \includegraphics[width=\columnwidth]{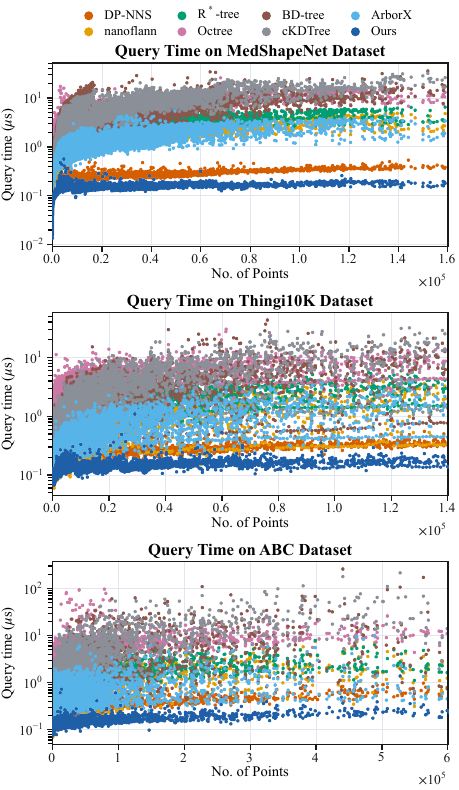}
    \caption{Comparison of average nearest-neighbor query times across the Thingi10K, ABC, and MedShapeNet datasets. Our method consistently outperforms existing baselines.}
    \label{fig:datasetquery}
\end{figure}

We further evaluate our method on the Thingi10K, ABC, and MedShapeNet datasets. As shown in Fig.~\ref{fig:datasetquery}, our method consistently outperforms existing baselines across all three datasets. These results demonstrate that our query speedup generalizes beyond the seven benchmark models to a broad range of surface point clouds.

\begin{figure}[t]
    \centering
    \includegraphics[width=\columnwidth]{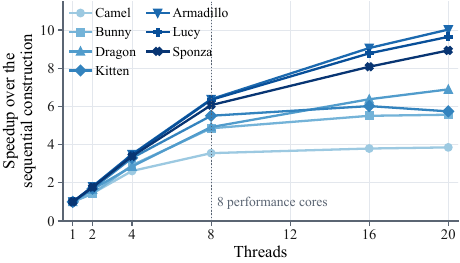}
    \caption{Construction speedup of our level-consistent parallel method over the sequential baseline across varying thread counts.}
    \label{fig:build_scaling}
\end{figure}

\subsection{Construction Time}
\label{sec:exp_construction}

Table~\ref{tab:construction} compares construction times across the seven benchmark models. The original DP-NNS, R*-tree, PCL octree, SciPy, and ANN implementations are evaluated sequentially as they do not support parallel construction, whereas the remaining methods are built in parallel. Our level-consistent parallel construction matches or outperforms CGAL's parallel insertion. Fig.~\ref{fig:build_scaling} illustrates how our construction speedup scales with the thread count, achieving a 4--10$\times$ speedup over the sequential baseline. Despite this acceleration, our construction remains 4--8$\times$ slower than building a nanoflann $k$-d tree.

\begin{figure*}[t]
    \centering
    \includegraphics[width=\textwidth]{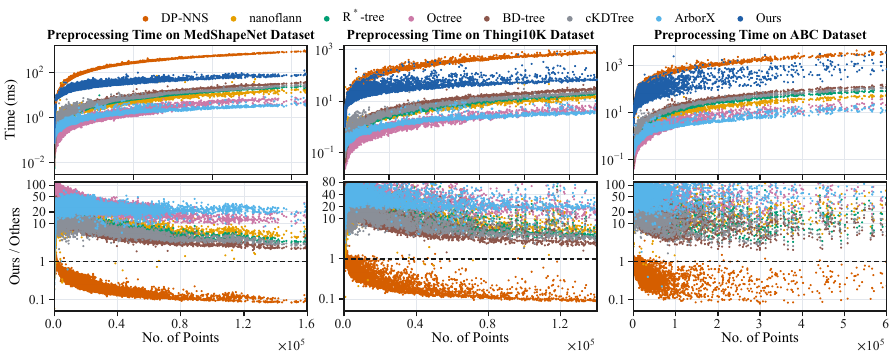}
    \caption{Construction times across the Thingi10K, ABC, and MedShapeNet datasets. Top: construction time. Bottom: ratio of our construction time to each competing method, where the dashed line ($1\times$) indicates equal time. Methods supporting parallel construction (nanoflann, ArborX, CGAL, and ours) run with 20 threads, while the rest run sequentially.}
    \label{fig:thingi_build}
\end{figure*}

Fig.~\ref{fig:thingi_build} extends this comparison to the Thingi10K, ABC, and MedShapeNet datasets, demonstrating substantial reductions in construction time compared to the original sequential implementation. Although our construction remains slower than that of tree-based indices, this significant reduction in preprocessing overhead relative to the sequential baseline substantially broadens the practical applicability of DP-NNS.

\begin{figure}[t]
    \centering
    \includegraphics[width=\columnwidth]{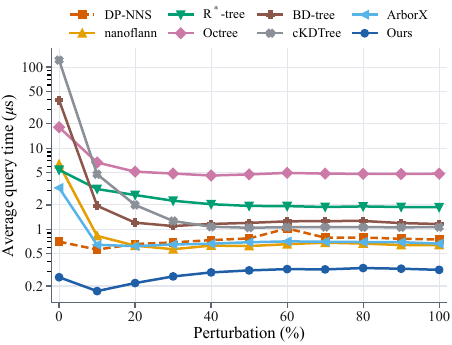}
    \caption{Average nearest-neighbor query time on the perturbed Lucy model, where each point is displaced along a random direction by a uniform distance up to a given percentage of the bounding-box diagonal. Our method achieves the fastest query times across all perturbation levels.}
    \label{fig}
\end{figure}

\subsection{Ablation Studies}
\label{sec:exp_ablation_extension}

\begin{table}[t]
    \centering
    \caption{Average nearest-neighbor query time ($\mu$s) with both data and query points sampled uniformly within the same 3D bounding box. Best and second-best results are \textbf{\underline{bold underlined}} and \underline{underlined}.}
    \label{tab}
    \setlength{\tabcolsep}{3.5pt}
    \footnotesize
    \begin{tabular}{l rrrrrr}
        \toprule
        Method & 10K & 50K & 100K & 200K & 500K & 1M \\
        \midrule
\uniformTableBody
        \bottomrule
    \end{tabular}
\end{table}

\paragraph{Data Distribution}
Our method targets a common graphics setting where data points lie on a 2D manifold in 3D space and queries span the surrounding volume. To test robustness beyond this assumption, we evaluate two non-manifold settings: (1) a fully volumetric setting with both data and query points sampled uniformly inside a 3D box (Table~\ref{tab}), and (2) a perturbed surface setting obtained by adding progressively larger uniform random displacements (proportional to the bounding-box diagonal) to the points of the Lucy model (Fig.~\ref{fig}). In both cases, our method consistently achieves the fastest query times across all tested distributions and noise levels.

\begin{table}[t]
    \centering
    \caption{Average nearest-neighbor query time ($\mu$s) with ($\checkmark$) and without ($\times$) grid-accelerated search initialization (G) and Morton query reordering (R).}
    \label{tab:grid_sort_ablation}
    \setlength{\tabcolsep}{2.2pt}
    \scriptsize
    \begin{tabular}{ccccccccc}
        \toprule
        G & R & Camel & Bunny & Dragon & Kitten & Armadillo & Lucy & Sponza \\
        \midrule
\gridSortAblationTableBody
        \bottomrule
    \end{tabular}
\end{table}

\begin{table}[t]
    \centering
    \caption{Effect of the grid resolution $G$ (number of cells per axis) on performance. Pre.: construction time (ms); Query: average nearest-neighbor query time ($\mu$s).}
    \label{tab:grid_resolution}
    \setlength{\tabcolsep}{1.6pt}
    \setlength{\aboverulesep}{0pt}\setlength{\belowrulesep}{0pt}\setlength{\extrarowheight}{1.2pt}
    \newcolumntype{?}{!{\color{black!35}\vrule width 0.4pt}}
    \scriptsize
    \begin{tabular}{c?cc?cc?cc?cc?cc}
        \toprule
        & \multicolumn{2}{c?}{$G=1$} & \multicolumn{2}{c?}{$G=16$} & \multicolumn{2}{c?}{$G=32$} & \multicolumn{2}{c?}{$G=64$} & \multicolumn{2}{c}{$G=128$} \\
        \cmidrule{2-3}\cmidrule{4-5}\cmidrule{6-7}\cmidrule{8-9}\cmidrule{10-11}
        Model & Pre. & Query & Pre. & Query & Pre. & Query & Pre. & Query & Pre. & Query \\
        \midrule
\gridResolutionTableBody
        \bottomrule
    \end{tabular}
\end{table}

\paragraph{Grid Initialization and Spatial Query Reordering}
To isolate the individual and combined contributions of grid-accelerated search initialization and spatial query reordering, we conduct a $2 \times 2$ ablation study across the seven benchmark models (Table~\ref{tab:grid_sort_ablation}). Compared to the unaccelerated baseline, enabling either strategy alone consistently reduces query time, with grid initialization yielding a more pronounced speedup than spatial query reordering. Combining both strategies achieves the lowest query time across all models, demonstrating that shorter search paths (via grid initialization) and improved cache locality (via spatial query reordering) provide complementary performance gains.

\begin{table}[t]
    \centering
    \caption{Average ICP registration time in seconds across four scenes. Best and second-best results are \textbf{\underline{bold underlined}} and \underline{underlined}.}
    \label{tab:icp}
    \setlength{\tabcolsep}{4pt}
    \footnotesize
    \begin{tabular}{l rrrr}
        \toprule
        Method & Living room 1 & Living room 2 & Office 1 & Office 2 \\
        \midrule
\appIcpTableBody
        \bottomrule
    \end{tabular}
\end{table}
\paragraph{Grid Resolution}
We further evaluate the impact of the grid resolution $G$ on construction and query times across four representative models (Table~\ref{tab:grid_resolution}). As $G$ increases, the construction time grows steadily due to the higher overhead of initializing grid cells. Conversely, the query time drops sharply at first and then plateaus at higher resolutions. Our default choice of $G$ strikes a favorable balance between construction overhead and query performance.

\subsection{Application to Point Cloud Registration}
\label{sec:application}

Point cloud registration is a fundamental task in computer graphics and computer vision. A widely used approach is Iterative Closest Point (ICP), which alternates between finding nearest-neighbor correspondences and estimating a rigid transformation. Because correspondence search is performed at every iteration and forms the primary computational bottleneck, accelerating this search backend is of substantial practical significance. To evaluate our method in this context, we integrate seven nearest-neighbor search backends into PCL's point-to-plane ICP implementation while keeping the remaining registration procedure unchanged. The experiments use adjacent fragment pairs from the four scenes of the Augmented ICL-NUIM dataset. For each pair, we perturb the ground-truth transformation with a $10^\circ$ rotation about a random axis and a $10\text{ cm}$ translation in a random direction to obtain the initial alignment.

Table~\ref{tab:icp} reports the average registration time per pair, including index construction and all ICP steps. Across all four scenes, our method achieves the fastest registration, with speedups of 1.7--1.8$\times$ over DP-NNS, the fastest competing method, and 18--24$\times$ over PCL's default $k$-d tree. By improving the efficiency of the complete pipeline through a change to its search backend, these results demonstrate the practical benefits of our method in accelerating real-world applications.

\section{Limitations and Future Work}
\label{sec:limitations}

Despite its empirical gains, our method has three main limitations that point to directions for future work. 
First, although our level-consistent parallel construction substantially lowers preprocessing overhead compared to sequential DP-NNS, building the Delaunay hierarchy remains slower than constructing tree-based indices (e.g., $k$-d trees). 
Second, our acceleration strategies are specifically tailored to exact single nearest-neighbor ($1$-NN) search and yield limited gains for general $k$-nearest-neighbor ($k$-NN) queries. 
Third, inherited from exact Delaunay triangulations, our approach is restricted to low-dimensional spaces (e.g., 3D), as construction complexity scales unfavorably with dimensionality. 
In future work, we plan to further reduce construction overhead, design specialized traversal mechanisms to accelerate $k$-NN search, 
and explore approximate Delaunay graph constructions to extend our framework to higher-dimensional spaces.

\section{Conclusion}
\label{sec:conclusion}

In this paper, we have presented an accelerated framework for exact nearest-neighbor search in 3D point clouds based on the Delaunay hierarchy (DP-NNS). 
To address the high preprocessing overhead of the original sequential structure, we developed a level-consistent parallel construction algorithm, analyzing its theoretical complexity and achieving substantial speedups across multi-core architectures.
To further boost query execution, we combined a uniform grid for search initialization with Morton query reordering, simultaneously shortening graph traversal paths and improving cache locality.

Extensive evaluations across classic benchmark models and three large-scale datasets—Thingi10K, ABC, and MedShapeNet—demonstrate that our method consistently achieves the fastest nearest-neighbor queries among state-of-the-art spatial indices, remaining robust across both surface-sampled manifolds and volumetric distributions. Furthermore, integrating our solver as a drop-in search backend into point-to-plane ICP registration yields significant end-to-end speedups, underscoring its practical value for real-world geometry processing pipelines.

\bibliographystyle{IEEEtran}
\bibliography{cvmbib}

@IEEEtranBSTCTL{IEEEexample:BSTcontrol,
  CTLuse_url = "no"
}

@article{dpnns,
  title={Efficient Nearest Neighbor Search Using Dynamic Programming},
  author={Peng-Fei Wang and Jiantao Song and Shi-Qing Xin and Shuang-Min Chen and Chang-He Tu and Wen-Ping Wang and Jia-Ye Wang},
  journal={IEEE Transactions on Pattern Analysis and Machine Intelligence},
  year={2026},
  volume={48},
  pages={999-1014},
  url={https://api.semanticscholar.org/CorpusID:272827847}
}

@article{manifoldknn,
  title={Manifold k-NN: Accelerated k-NN Queries for Manifold Point Clouds},
  author={Peng-Fei Wang and Qing Guo and Hai-Sen Zhao and Shi-Qing Xin and Shuang-Min Chen and Chang-He Tu and Wen-Ping Wang},
  journal={ACM Transactions on Graphics (TOG)},
  year={2026},
  volume={45},
  pages={1 - 12},
  url={https://api.semanticscholar.org/CorpusID:287950612}
}

@incollection{cgal:hs-chdt3-24a
, author =  "Susan Hert and Michael Seel"
, title =   "{dD} Convex Hulls and Delaunay Triangulations"
, publisher =  "{CGAL Editorial Board}"
, edition =     "{5.6.1}"
, booktitle =   "{CGAL} User and Reference Manual"
, url = "https://doc.cgal.org/5.6.1/Manual/packages.html#PkgConvexHullD"
, year =        2024
}

@article{virtanen2020scipy,
  title={SciPy 1.0: fundamental algorithms for scientific computing in Python},
  author={Pauli Virtanen and Ralf Gommers and Travis E. Oliphant and Matt Haberland and Tyler Reddy and David Cournapeau and Evgeni Burovski and Pearu Peterson and Warren Weckesser and Jonathan Bright and St{\'e}fan J. van der Walt and Matthew Brett and Joshua Wilson and K. Jarrod Millman and Nikolay Mayorov and Andrew R. J. Nelson and Eric Jones and Robert Kern and Eric B. Larson and Clifton Carey and Ilhan Polat and Yu Feng and Eric W. Moore and Jacob T. VanderPlas and Denis Laxalde and Josef Perktold and Robert Cimrman and Ian Daniel Henriksen and E. A. Quintero and Charles R. Harris and Anne M. Archibald and Ant{\^o}nio H. Ribeiro and Fabian Pedregosa and Paul van Mulbregt and Aditya Vijaykumar and Alessandro Pietro Bardelli and Alex Rothberg and Andreas Hilboll and Andre Kloeckner and Anthony M. Scopatz and Antony R. Lee and Ariel Rokem and C. Nathan Woods and Chad Fulton and Charles Masson and Christian H{\"a}ggstr{\"o}m and Clark Fitzgerald and David A. Nicholson and David R. Hagen and Dmitrii V. Pasechnik and Emanuele Olivetti and Eric Martin and Eric Wieser and Fabrice Silva and Felix Lenders and Florian Wilhelm and Gideon Young and Gavin A. Price and Gert-Ludwig Ingold and Gregory E. Allen and Gregory R. Lee and Herv{\'e} Audren and Irvin Probst and J. P. Dietrich and Jacob Silterra and James T. Webber and Janko Slavi{\v{c}}   and Joel Nothman and Johannes Buchner and Johannes Kulick and Johannes L. Sch{\"o}nberger and Jos{\'e} Vin{\'i}cius de Miranda Cardoso and Joscha Reimer and Joseph E. Harrington and Juan Luis Cano Rodr{\'i}guez and Juan Nunez-Iglesias and Justin Kuczynski and Kevin Lee Tritz and Martin Thoma and Matt Newville and Matthias K{\"u}mmerer and Maximilian Bolingbroke and Michael Tartre and Mikhail Pak and Nathaniel J. Smith and Nikolai Nowaczyk and Nikolay Shebanov and Oleksandr Pavlyk and Per Andreas Brodtkorb and Perry Lee and Robert T. McGibbon and Roman Feldbauer and Sam Lewis and Sam Tygier and Scott Sievert and Sebastiano Vigna and Stefan Peterson and Surhud More and Tadeusz Pudlik and Taku Oshima and Thomas J. Pingel and Thomas P. Robitaille and Thomas Spura and Thouis Raymond Jones and Tim Cera and Tim Leslie and Tiziano Zito and Tom Krauss and U. Upadhyay and Yaroslav O. Halchenko and Y. V{\'a}zquez-Baeza},
  journal={Nature Methods},
  year={2019},
  volume={17},
  pages={261 - 272},
  url={https://api.semanticscholar.org/CorpusID:198229805}
}

@misc{nanoflann,
author = {Blanco-Claraco, Jos{\'e} Luis and others},
title = {nanoflann: a {C++11} Header-Only Library for Nearest-Neighbor Search},
year = {2025},
howpublished = {Version 1.7.1, \url{https://github.com/jlblancoc/nanoflann}}
}

@manual{boostgeometry,
author = {Gehrels, Barend and Lalande, Bruno and Loskot, Mateusz and Wulkiewicz, Adam},
title = {Boost.Geometry Index Reference},
organization = {Boost C++ Libraries},
edition = {1.84},
year = {2023},
url = {https://www.boost.org/doc/libs/1_84_0/libs/geometry/doc/html/geometry/reference/spatial_indexes.html}
}

@article{pcl,
  title={3D is here: Point Cloud Library (PCL)},
  author={Radu Bogdan Rusu and Steve B. Cousins},
  journal={2011 IEEE International Conference on Robotics and Automation},
  year={2011},
  pages={1-4},
  url={https://api.semanticscholar.org/CorpusID:206849822}
}

@inproceedings{annlibrary,
  title={ANN: library for approximate nearest neighbor searching},
  author={Sunil Arya and David M. Mount},
  year={1998},
  url={https://api.semanticscholar.org/CorpusID:117479624}
}

@article{arborx2,
  title={The ArborX Library: Version 2.0},
  author={Andrey Prokopenko and Daniel Arndt and Damien Lebrun-Grandi{\'e} and Bruno Turcksin},
  journal={ACM Transactions on Mathematical Software},
  year={2025},
  volume={51},
  pages={1 - 10},
  url={https://api.semanticscholar.org/CorpusID:280401698}
}

@article{bentley1975,
  title={Multidimensional binary search trees used for associative searching},
  author={Jon Louis Bentley},
  journal={Commun. ACM},
  year={1975},
  volume={18},
  pages={509-517},
  url={https://api.semanticscholar.org/CorpusID:13091446}
}

@article{friedman1977,
  title={An Algorithm for Finding Best Matches in Logarithmic Expected Time},
  author={Jerome H. Friedman and Jon Louis Bentley and Raphael A. Finkel},
  journal={ACM Transactions on Mathematical Software (TOMS)},
  year={1977},
  volume={3},
  pages={209 - 226},
  url={https://api.semanticscholar.org/CorpusID:10811510}
}

@article{meagher1982,
  title={Geometric modeling using octree encoding},
  author={Donald J. Meagher},
  journal={Comput. Graph. Image Process.},
  year={1982},
  volume={19},
  pages={85},
  url={https://api.semanticscholar.org/CorpusID:205045003}
}

@inproceedings{omohundro1989,
  title={Five Balltree Construction Algorithms},
  author={Stephen M. Omohundro},
  year={2009},
  url={https://api.semanticscholar.org/CorpusID:61067117}
}

@inproceedings{yianilos1993,
  title={Data structures and algorithms for nearest neighbor search in general metric spaces},
  author={Peter N. Yianilos},
  booktitle={ACM-SIAM Symposium on Discrete Algorithms},
  year={1993},
  url={https://api.semanticscholar.org/CorpusID:2871061}
}

@article{bowyer1981,
  title={Computing Dirichlet Tessellations},
  author={Adrian Bowyer},
  journal={Comput. J.},
  year={1981},
  volume={24},
  pages={162-166},
  url={https://api.semanticscholar.org/CorpusID:44282750}
}

@article{watson1981,
  title={Computing the n-Dimensional Delaunay Tesselation with Application to Voronoi Polytopes},
  author={D. F. Watson},
  journal={Comput. J.},
  year={1981},
  volume={24},
  pages={167-172},
  url={https://api.semanticscholar.org/CorpusID:27550605}
}

@inproceedings{amenta2003brio,
  title={Incremental constructions con BRIO},
  author={Nina Amenta and Sung-Hee Choi and G{\"u}nter Rote},
  booktitle={SCG '03},
  year={2003},
  url={https://api.semanticscholar.org/CorpusID:3992843}
}

@inproceedings{batista2010,
  title={Parallel geometric algorithms for multi-core computers},
  author={Vicente H. F. Batista and David L. Millman and Sylvain Pion and John Victor Singler},
  booktitle={SCG '09},
  year={2009},
  url={https://api.semanticscholar.org/CorpusID:15797117}
}

@inproceedings{blelloch2012,
  title={Internally deterministic parallel algorithms can be fast},
  author={Guy E. Blelloch and Jeremy T. Fineman and Phillip B. Gibbons and Julian Shun},
  booktitle={ACM SIGPLAN Symposium on Principles \& Practice of Parallel Programming},
  year={2012},
  url={https://api.semanticscholar.org/CorpusID:7201965}
}

@article{blelloch2016,
  title={Parallelism in Randomized Incremental Algorithms},
  author={Guy E. Blelloch and Yan Gu and Julian Shun and Yi-Han Sun},
  journal={Journal of the ACM (JACM)},
  year={2018},
  volume={67},
  pages={1 - 27},
  url={https://api.semanticscholar.org/CorpusID:12349988}
}

@article{Ost2021NeuralPL,
  title={Neural Point Light Fields},
  author={Julian Ost and Issam Hadj Laradji and Alejandro Newell and Yuval Bahat and Felix Heide},
  journal={2022 IEEE/CVF Conference on Computer Vision and Pattern Recognition (CVPR)},
  year={2021},
  pages={18398-18408},
  url={https://api.semanticscholar.org/CorpusID:244799553}
}

@article{Chang2023PointersectNR,
  title={Pointersect: Neural Rendering with Cloud-Ray Intersection},
  author={Jen-Hao Rick Chang and Wei-Yu Chen and Anurag Ranjan and Kwang Moo Yi and Oncel Tuzel},
  journal={2023 IEEE/CVF Conference on Computer Vision and Pattern Recognition (CVPR)},
  year={2023},
  pages={8359-8369},
  url={https://api.semanticscholar.org/CorpusID:258309623}
}

@article{Chen2023NeuralEditorEN,
  title={NeuralEditor: Editing Neural Radiance Fields via Manipulating Point Clouds},
  author={Junkun Chen and Jipeng Lyu and Yu-Xiong Wang},
  journal={2023 IEEE/CVF Conference on Computer Vision and Pattern Recognition (CVPR)},
  year={2023},
  pages={12439-12448},
  url={https://api.semanticscholar.org/CorpusID:258479756}
}

@article{Wang2021NeuralIMLSSI,
  title={Neural-IMLS: Self-Supervised Implicit Moving Least-Squares Network for Surface Reconstruction},
  author={Zixiong Wang and Peng-Fei Wang and Peng Wang and Qiujie Dong and Junjie Gao and Shuang-Min Chen and Shi-Qing Xin and Chang-He Tu and Wen-Ping Wang},
  journal={IEEE Transactions on Visualization and Computer Graphics},
  year={2021},
  volume={30},
  pages={5018-5033},
  url={https://api.semanticscholar.org/CorpusID:253420849}
}

@inproceedings{10.1145/602259.602266,
  title={R-trees: a dynamic index structure for spatial searching},
  author={Antonin Guttman},
  booktitle={ACM SIGMOD Conference},
  year={1984},
  url={https://api.semanticscholar.org/CorpusID:876601}
}

@inproceedings{rstarTREE,
  title={The R*-tree: an efficient and robust access method for points and rectangles},
  author={Norbert Beckmann and Hans-Peter Kriegel and Ralf Schneider and Bernhard Seeger},
  booktitle={ACM SIGMOD Conference},
  year={1990},
  url={https://api.semanticscholar.org/CorpusID:11567855}
}

@article{Gao2025Parallel3D,
  title={Parallel 3D Delaunay Triangulation on the GPU},
  author={Wu Gao and Renjie Chen},
  journal={Comput. Aided Des.},
  year={2025},
  volume={189},
  pages={103933},
  url={https://api.semanticscholar.org/CorpusID:280736740}
}

@article{Gao2026LocalDB,
  title={Local DeWall: Billion-scale 3D Delaunay triangulation on a single GPU},
  author={Wu Gao and Renjie Chen},
  journal={Comput. Aided Des.},
  year={2026},
  volume={199},
  pages={104107},
  url={https://api.semanticscholar.org/CorpusID:289252550}
}

@InProceedings{Koch_2019_CVPR,
author = {Koch, Sebastian and Matveev, Albert and Jiang, Zhongshi and Williams, Francis and Artemov, Alexey and Burnaev, Evgeny and Alexa, Marc and Zorin, Denis and Panozzo, Daniele},
title = {ABC: A Big CAD Model Dataset for Geometric Deep Learning},
booktitle = {Proceedings of the IEEE/CVF Conference on Computer Vision and Pattern Recognition (CVPR)},
month = {June},
year = {2019}
}

@article{medshapenet,
  TITLE = {{MedShapeNet -- A Large-Scale Dataset of 3D Medical Shapes for Computer Vision}},
  AUTHOR = {Li, Jianning and Zhou, Zongwei and Yang, Jiancheng and Pepe, Antonio and Gsaxner, Christina and Luijten, Gijs and Qu, Chongyu and Zhang, Tiezheng and Chen, Xiaoxi and Li, Wenxuan and Wodzinski, Marek and Friedrich, Paul and Xie, Kangxian and Jin, Yuan and Ambigapathy, Narmada and Nasca, Enrico and Solak, Naida and Melito, Gian Marco and Vu, Viet Duc and Memon, Afaque R. and Schlachta, Christopher and de Ribaupierre, Sandrine and Patel, Rajnikant and Eagleson, Roy and Chen, Xiaojun and M{\"a}chler, Heinrich and Kirschke, Jan Stefan and de la Rosa, Ezequiel and Christ, Patrick Ferdinand and Li, Hongwei Bran and Ellis, David G. and Aizenberg, Michele R. and Gatidis, Sergios and K{\"u}stner, Thomas and Shusharina, Nadya and Heller, Nicholas and Andrearczyk, Vincent and Depeursinge, Adrien and Hatt, Mathieu and Sekuboyina, Anjany and L{\"o}ffler, Maximilian and Liebl, Hans and Dorent, Reuben and Vercauteren, Tom and Shapey, Jonathan and Kujawa, Aaron and Cornelissen, Stefan and Langenhuizen, Patrick and Ben-Hamadou, Achraf and Rekik, Ahmed and Pujades, Sergi and Boyer, Edmond and Bolelli, Federico and Grana, Costantino and Lumetti, Luca and Salehi, Hamidreza and Ma, Jun and Zhang, Yao and Gharleghi, Ramtin and Beier, Susann and Sowmya, Arcot and Garza-Villarreal, Eduardo A. and Balducci, Thania and Angeles-Valdez, Diego and Souza, Roberto and Rittner, Leticia and Frayne, Richard and Ji, Yuanfeng and Ferrari, Vincenzo and Chatterjee, Soumick and Dubost, Florian and Schreiber, Stefanie and Mattern, Hendrik and Speck, Oliver and Haehn, Daniel and John, Christoph and N{\"u}rnberger, Andreas and Pedrosa, Jo{\~a}o and Ferreira, Carlos and Aresta, Guilherme and Cunha, Ant{\'o}nio and Campilho, Aur{\'e}lio and Suter, Yannick and Garcia, Jose and Lalande, Alain and Vandenbossche, Vicky and van Oevelen, Aline and Duquesne, Kate and Mekhzoum, Hamza and Vandemeulebroucke, Jef and Audenaert, Emmanuel and Krebs, Claudia and van Leeuwen, Timo and Vereecke, Evie and Heidemeyer, Hauke and R{\"o}hrig, Rainer and H{\"o}lzle, Frank and Badeli, Vahid and Krieger, Kathrin and Gunzer, Matthias and Chen, Jianxu and van Meegdenburg, Timo and Dada, Amin and Balzer, Miriam and Fragemann, Jana and Jonske, Frederic and Rempe, Moritz and Malorodov, Stanislav and Bahnsen, Fin H. and Seibold, Constantin and Jaus, Alexander and Marinov, Zdravko and Jaeger, Paul F. and Stiefelhagen, Rainer and Santos, Ana Sofia and Lindo, Mariana and Ferreira, Andr{\'e} and Alves, Victor and Kamp, Michael and Abourayya, Amr and Nensa, Felix and H{\"o}rst, Fabian and Brehmer, Alexander and Heine, Lukas and Hanusrichter, Yannik and We{\ss}ling, Martin and Dudda, Marcel and Podleska, Lars E. and Fink, Matthias A. and Keyl, Julius and Tserpes, Konstantinos and Kim, Moon-Sung and Elhabian, Shireen and Lamecker, Hans and Zuki{\'c}, D{\v z}enan and Paniagua, Beatriz and Wachinger, Christian and Urschler, Martin and Duong, Luc and Wasserthal, Jakob and Hoyer, Peter F. and Basu, Oliver and Maal, Thomas and Witjes, Max J. H. and Schiele, Gregor and Chang, Ti-Chiun and Ahmadi, Seyed-Ahmad and Luo, Ping and Menze, Bjoern and Reyes, Mauricio and Deserno, Thomas M. and Davatzikos, Christos and Puladi, Behrus and Fua, Pascal and Yuille, Alan L. and Kleesiek, Jens and Egger, Jan},
  URL = {https://inria.hal.science/hal-04397470},
  JOURNAL = {{Biomedical Engineering = Biomedizinische Technik}},
  PUBLISHER = {{De Gruyter}},
  VOLUME = {70},
  NUMBER = {1},
  PAGES = {71-90},
  YEAR = {2025},
  MONTH = Feb,
  DOI = {10.1515/bmt-2024-0396},
  HAL_ID = {hal-04397470},
  HAL_VERSION = {v1},
}

@article{thingi10k,
  title={Thingi10K: A Dataset of 10,000 3D-Printing Models},
  author={Zhou, Qingnan and Jacobson, Alec},
  journal={arXiv preprint arXiv:1605.04797},
  year={2016}
}

@inproceedings{blandford2006compact,
  title={Engineering a compact parallel delaunay algorithm in 3D},
  author={Daniel K. Blandford and Guy E. Blelloch and Clemens Kadow},
  booktitle={SCG '06},
  year={2006},
  url={https://api.semanticscholar.org/CorpusID:2663966}
}

@inproceedings{foteinos2011dynamic,
  title={Dynamic Parallel 3D Delaunay Triangulation},
  author={Panagiotis A. Foteinos and Nikos Chrisochoides},
  booktitle={International Meshing Roundtable Conference},
  year={2011},
  url={https://api.semanticscholar.org/CorpusID:11665944}
}

@article{cao2014gdel3d,
  title={A GPU accelerated algorithm for 3D Delaunay triangulation},
  author={Thanh-Tung Cao and Ashwin Nanjappa and Mingcen Gao and Tiow-Seng Tan},
  journal={Proceedings of the 18th meeting of the ACM SIGGRAPH Symposium on Interactive 3D Graphics and Games},
  year={2014},
  url={https://api.semanticscholar.org/CorpusID:3347902}
}

@article{marot2019onemachine,
  title={One machine, one minute, three billion tetrahedra},
  author={C{\'e}lestin Marot and Jeanne Pellerin and Jean-François Remacle},
  journal={International Journal for Numerical Methods in Engineering},
  year={2018},
  volume={117},
  pages={967 - 990},
  url={https://api.semanticscholar.org/CorpusID:43929301}
}

@article{guibas1992randomized,
  title={Randomized incremental construction of Delaunay and Voronoi diagrams},
  author={Leonidas J. Guibas and Donald Ervin Knuth and Micha Sharir},
  journal={Algorithmica},
  year={1990},
  volume={7},
  pages={381-413},
  url={https://api.semanticscholar.org/CorpusID:3770886}
}

@article{boissonnat1993delaunaytree,
  title={On the Randomized Construction of the Delaunay Tree},
  author={Jean-Daniel Boissonnat and Monique Teillaud},
  journal={Theor. Comput. Sci.},
  year={1993},
  volume={112},
  pages={339-354},
  url={https://api.semanticscholar.org/CorpusID:167301}
}

@article{devillers2002hierarchy,
  title={The Delaunay Hierarchy},
  author={Olivier Devillers},
  journal={Int. J. Found. Comput. Sci.},
  year={2002},
  volume={13},
  pages={163-180},
  url={https://api.semanticscholar.org/CorpusID:14268767}
}

@article{hoppe1992surface,
  title={Surface reconstruction from unorganized points},
  author={Hugues Hoppe and Tony DeRose and Tom Duchamp and John Alan McDonald and Werner Stuetzle},
  journal={Proceedings of the 19th annual conference on Computer graphics and interactive techniques},
  year={1992},
  url={https://api.semanticscholar.org/CorpusID:1122241}
}

@article{ma2021neuralpull,
  title={Neural-Pull: Learning Signed Distance Functions from Point Clouds by Learning to Pull Space onto Surfaces},
  author={Baorui Ma and Zhi-Zhong Han and Yu-Shen Liu and Matthias Zwicker},
  journal={ArXiv},
  year={2020},
  volume={abs/2011.13495},
  url={https://api.semanticscholar.org/CorpusID:227209266}
}

@article{zhao2020song,
  title={SONG: Approximate Nearest Neighbor Search on GPU},
  author={Wei-Jie Zhao and Shu-Long Tan and Ping Li},
  journal={2020 IEEE 36th International Conference on Data Engineering (ICDE)},
  year={2020},
  pages={1033-1044},
  url={https://api.semanticscholar.org/CorpusID:218906516}
}

@article{groh2023ggnn,
  title={GGNN: Graph-Based GPU Nearest Neighbor Search},
  author={Fabian Groh and Lukas Ruppert and Patrick Wieschollek and Hendrik P. A. Lensch},
  journal={IEEE Transactions on Big Data},
  year={2019},
  volume={9},
  pages={267-279},
  url={https://api.semanticscholar.org/CorpusID:208548477}
}

@article{ootomo2024cagra,
  title={CAGRA: Highly Parallel Graph Construction and Approximate Nearest Neighbor Search for GPUs},
  author={Hiroyuki Ootomo and Akira Naruse and Corey J. Nolet and Ray Wang and Tamas B. Feh{\'e}r and Y. Wang},
  journal={2024 IEEE 40th International Conference on Data Engineering (ICDE)},
  year={2023},
  pages={4236-4247},
  url={https://api.semanticscholar.org/CorpusID:261276680}
}

@article{devillers2002walking,
  title={Walking in a triangulation},
  author={Olivier Devillers and Sylvain Pion and Monique Teillaud},
  journal={Int. J. Found. Comput. Sci.},
  year={2001},
  volume={13},
  pages={181-199},
  url={https://api.semanticscholar.org/CorpusID:15112447}
}

@article{bose2004online,
  title={Online Routing in Triangulations},
  author={Prosenjit Bose and Pat Morin},
  journal={SIAM J. Comput.},
  year={1999},
  volume={33},
  pages={937-951},
  url={https://api.semanticscholar.org/CorpusID:755945}
}

@article{malkov2020hnsw,
  title={Efficient and Robust Approximate Nearest Neighbor Search Using Hierarchical Navigable Small World Graphs},
  author={Yury Malkov and D. A. Yashunin},
  journal={IEEE Transactions on Pattern Analysis and Machine Intelligence},
  year={2016},
  volume={42},
  pages={824-836},
  url={https://api.semanticscholar.org/CorpusID:8915893}
}

@article{kohout2005parallel,
  title={Parallel Delaunay triangulation in E2 and E3 for computers with shared memory},
  author={Josef Kohout and Ivana Kolingerov{\'a} and Jir{\'i} Z{\'a}ra},
  journal={Parallel Comput.},
  year={2005},
  volume={31},
  pages={491-522},
  url={https://api.semanticscholar.org/CorpusID:1086657}
}

@article{devillers2011perturbations,
  title={Perturbations for Delaunay and weighted Delaunay 3D triangulations},
  author={Olivier Devillers and Monique Teillaud},
  journal={Comput. Geom.},
  year={2011},
  volume={44},
  pages={160-168},
  url={https://api.semanticscholar.org/CorpusID:1898489}
}

@article{bentley1980optimal,
  title={Optimal Expected-Time Algorithms for Closest Point Problems},
  author={Jon Louis Bentley and Bruce W. Weide and Andrew Chi-Chih Yao},
  journal={ACM Transactions on Mathematical Software (TOMS)},
  year={1980},
  volume={6},
  pages={563 - 580},
  url={https://api.semanticscholar.org/CorpusID:17238717}
}

@article{mei2018degreedistributiondelaunaytriangulations,
  title={Degree Distribution of Delaunay Triangulations},
  author={Gang Mei and Neng-Xiong Xu and Salvatore Cuomo},
  journal={arXiv: Physics and Society},
  year={2018},
  url={https://api.semanticscholar.org/CorpusID:76656643}
}

\begin{IEEEbiography}[{\includegraphics[width=1in,height=1.25in,clip,keepaspectratio]{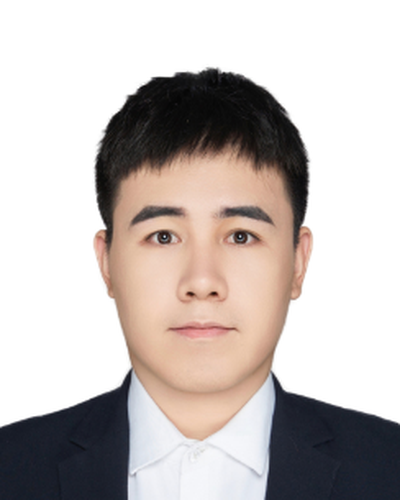}}]{Pengfei Wang}
is currently working toward the PhD degree in computer science with Shandong University. His research interests include computational geometry, computer graphics, and computer-aided design.
\end{IEEEbiography}

\begin{IEEEbiography}[{\includegraphics[width=1in,height=1.25in,clip,keepaspectratio]{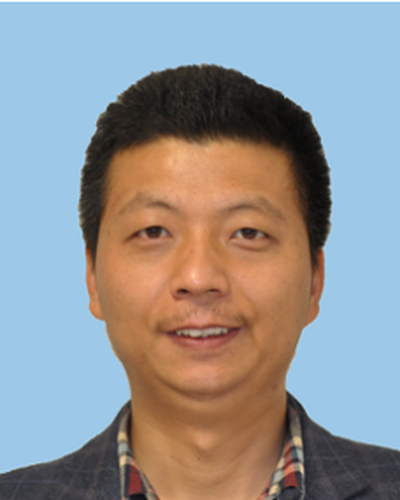}}]{Shiqing Xin}
is an associate professor at the School of Computer Science and Technology, Shandong University. He received his PhD degree in applied mathematics from Zhejiang University. His research focuses on geometric calculation, geometric modeling, and scene understanding.
\end{IEEEbiography}

\begin{IEEEbiography}[{\includegraphics[width=1in,height=1.25in,clip,keepaspectratio]{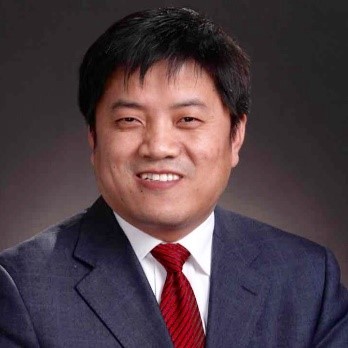}}]{Changhe Tu}
received the BSc, MEng, and PhD degrees from Shandong University, China, in 1990, 1993, and 2003, respectively. He is currently a professor with the School of Computer Science and Technology, Shandong University, China. He currently leads the CG-VIS Group, Shandong University. He has authored or coauthored more than 100 papers in international journals and conferences. His research interests include computer graphics, 3D vision, and computer-aided geometric design.
\end{IEEEbiography}

\end{document}